# OFF-state trapping phenomena in GaN HEMTs: interplay between gate trapping, acceptor ionization and positive charge redistribution

E. Canato[a], M. Meneghini[a], C. De Santi[a], F. Masin[a], A. Stockman[b], P. Moens[b], E. Zanoni[a], G. Meneghesso[a]

[a] *Department of Information Engineering, University of Padova, Italy*
[b] *ON Semiconductor, Oudenaarde, Belgium*

***Abstract*—We present an extensive analysis of the trapping processes induced by drain bias stress in AlGaN/GaN high-electron-mobility transistors (HEMTs) with p-GaN gate. We demonstrate that: (i) with increasing drain stress, pulsed I-V and $V_{TH}$ measurements shown an initial positive $V_{TH}$ variation and an increase in $R_{ON}$ then, for drain voltages >100 V, $V_{TH}$ is stable and the $R_{ON}$ shows a partial recovery. (ii) At moderate voltages, $V_{TH}$ instability is related to trapping at the gate stack, due to residual negative charge left behind by the holes that leave the p-GaN layer through the Schottky gate contact and/or to trapping at the barrier. At higher voltages, we demonstrate the interplay of two trapping processes by C-V and pulsed drain current analysis: (iii) a fast storage of positive charge, accumulated near the buffer/SRL interface, not strongly thermally activated, dominating at higher voltages; (iv) a slower negative charge storage, thermally activated with activation energies for trapping and de-trapping equal to ~0.6 eV and ~0.4-0.8 eV, respectively.**

## I. Introduction

GaN-based high electron mobility transistors (HEMTs) are expected to play a major role in next generation power converters, thanks to the low ON-resistance $R_{ON}$, fast switching speed, and high-temperature operation capability [1-5].

AlGaN/GaN HEMTs are intrinsically depletion-mode transistors with excellent performance, owing to the inherent high sheet carrier density at AlGaN/GaN hetero-interface caused by the material's unique polarization-induced charges; however, for low static power dissipation and safety in high power application, normally-off operation is of critical importance. In order to obtain an enhancement mode device, a p-type doped GaN layer is introduced under the gate metal [6-11]. To reduce gate leakage, the metal/p-GaN interface is typically of Schottky type.

Recently reports investigated the effects of drain bias stress on the threshold voltage $V_{TH}$ and ON-resistance in p-GaN gate HEMTs. Specifically, it was observed that a drain bias stress may induce a $V_{TH}$ shift of the p-GaN gate HEMT, ascribed to negative charge storage in the floating p-GaN layer [12]. Efthymiou et al. attributed the $V_{TH}$ instability due to off-state drain stress to the ionization of Mg acceptors traps followed by hole depletion in the AlGaN region below the p-GaN gate [13]. Other reports [14-18] proposed that trapping in off-state originates from the ionization of buffer acceptors ($C_N$ atoms), that results in a partial depletion of the channel. Despite the importance of these two processes, no paper described the interplay of these two mechanisms and the related trapping/de-trapping dynamics. Thus, it is not clear which of these processes dominates, as a function of stress time and applied voltage.

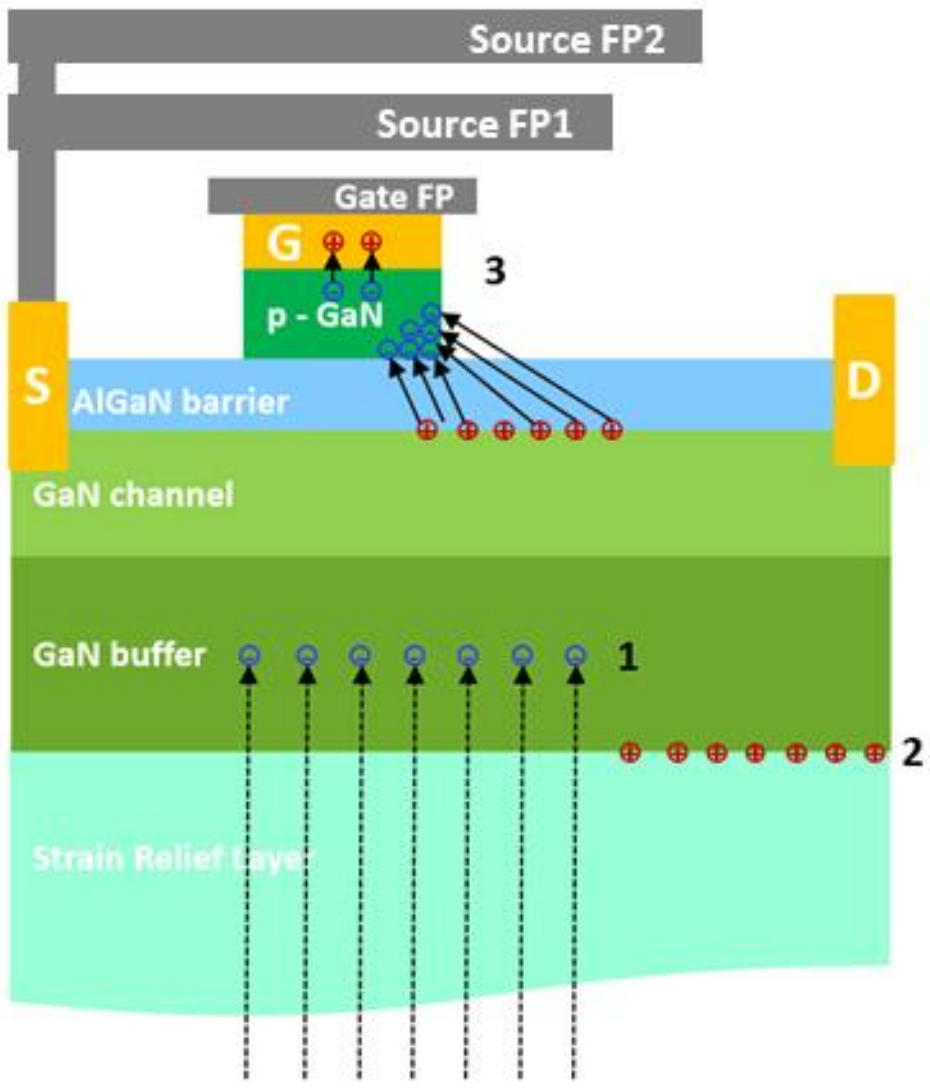

Fig. 1. Schematic representation of the mechanisms responsible for threshold voltage and ON-resistance variation. 1. ionization of carbon acceptors in the buffer; 2. storage of positive charge in the buffer; 3. storage of negative charges in the p-GaN layer.

The aim of this paper is to present a detailed study of the trapping processes induced by the off-state bias stress in p-GaN gate HEMT by means of different measurement techniques in order to distinguish the different competing mechanisms described above. For the first time, we present a detailed investigation of the interplay of trapping at the gate stack and in the buffer layer of GaN HEMTs, by identifying three dominant mechanisms: (i) ionization of carbon acceptors in the buffer; (ii) storage of positive charge in the buffer; (iii) storage of negative charges in the p-GaN layer. These mechanisms are described schematically in Figure 1. In addition, we describe the related dependence on voltage, temperature and time.

## II. Device Description and Measurements

The devices under test (DUT) in this paper are 650V rated normally-off lateral AlGaN/GaN-on-Si HEMT based on p-gate technology. The gate region was capped with a p-type doped GaN layer, on which a Schottky contact was formed. Field plates were formed in the region between the gate and the drain: one field plate was connected to the gate, and two field plates were connected to the source. PCM devices with 0.2-mm gate width and powerbars with 69mm-gate width were fabricated on the same wafer. DC characteristics were measured for PCM. The maximum average drain current at $V_{GS}$ = 6 V and $V_{DS}$ = 6V is 0.28 A/mm. The $V_{TH}$ is defined as the $V_{GS}$ at which the $I_D$ reaches 1μA/mm for $V_{DS}$ = 5 V and the mean value is 2.1 V.

## III. Measurements and Discussion

### *A. Pulsed characterization*

Initially the drain induced $V_{TH}$ and $R_{ON}$ instability is investigated by using pulsed I-V measurements in off-state conditions. Within one cycle, the DUT is first stressed in the OFF-state with $V_{GSQ}$ = 0 V and with a drain bias varying from 0 V up to 200 V, and then switched to the ON-state to measure the dynamic $V_{TH}$ and $R_{ON}$. The pulse width (duration of the ON-state measure phase) is 40 μs and the stress period is 4 ms. The results in Figure 2 (a) and (c) indicate an initial negative charge trapping until 70 V – 100 V that causes a positive $V_{TH}$ variation (Fig. 2 (b)) and an increase in $R_{ON}$ (Fig. 2 (d)). Then, for higher drain voltages (>100 V) $V_{TH}$ is stable (Fig. 2 (b)) and the $R_{ON}$ shows a partial recovery (Fig. 2 (d)). The non-monotonic variation of $R_{ON}$ and the positive $V_{TH}$ shift are ascribed to different mechanisms [12, 15]. Specifically, the non-monotonic trend of the $R_{ON}$ could be explained by the interplay between buffer trapping (ionization of $C_N$ acceptors, **mechanism 1**, taking place at moderate voltages <100 V) and the generation of positive charge in the buffer, (**mechanism 2**, taking place at higher voltages, >100 V), consistently with previous reports [15].

On the other hand, the positive $V_{TH}$ shift is ascribed to trapping of negative charge in the gate stack (**mechanism 3**), as preliminarily proposed in [12]. Specifically, when the device is in the off-state, holes may leave the p-GaN layer through the Schottky junction, thus leaving the ionized acceptors behind. When the drain bias drops, a portion of the negative charges remains stored in the p-GaN layer, since the Schottky junction is reversed biased and charge redistribution is a relatively slow process, resulting in a positive $V_{TH}$ shift [12]. An alternative explanation could be the trapping of electrons at the barrier [13].

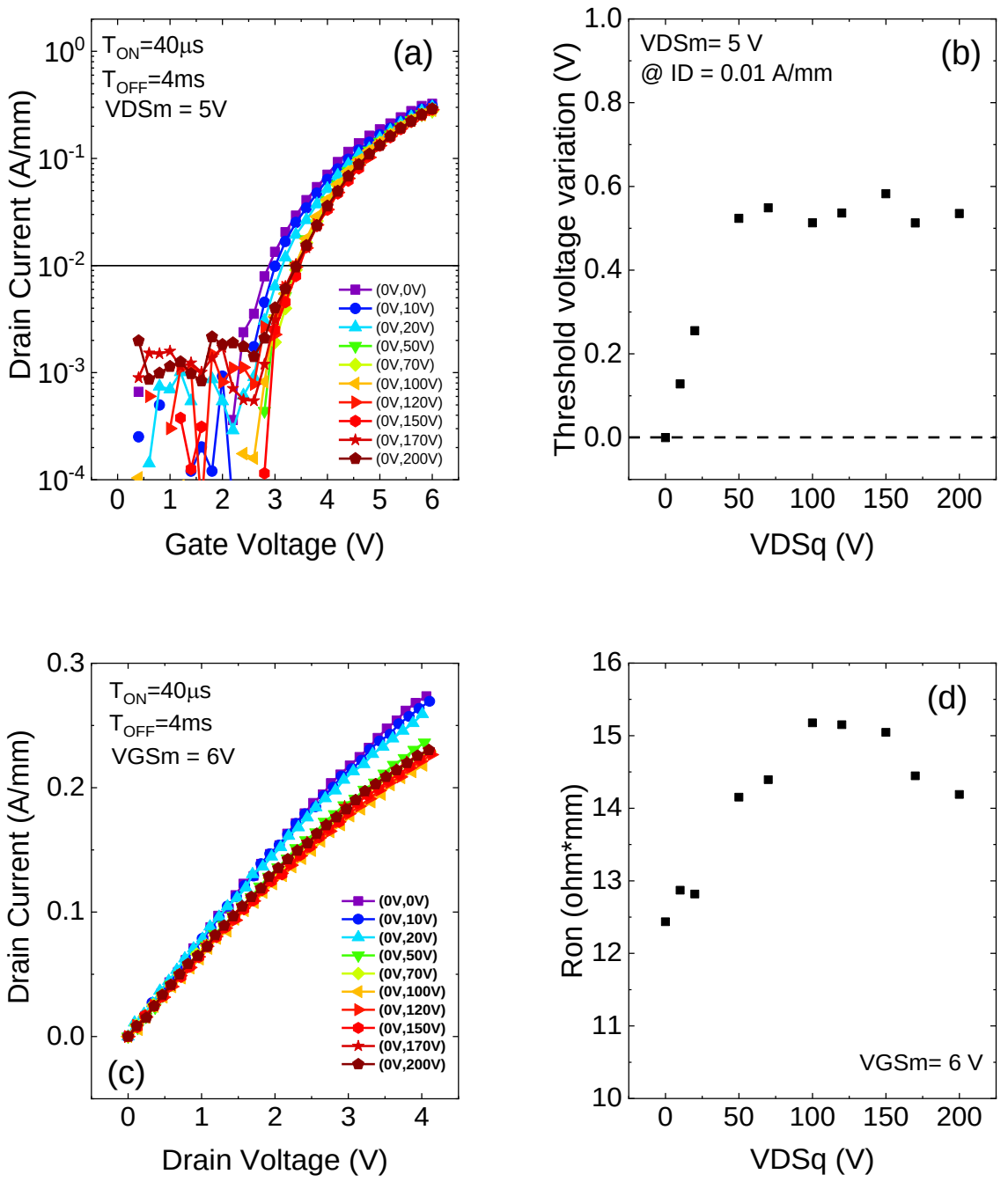

Fig. 2. Pulsed I-V measurements in off-state conditions: (a) $I_DV_G$ characteristic; (b) threshold voltage variation as a function of off-state drain stress; (c) $I_DV_D$ characteristic; (d) on resistance as a function of off-state drain stress.

### *B. High voltage C-V measurement*

The study of the capacitance-voltage (C-V) characteristics of the p-GaN gate HEMT can provide information on which region is affected by the trapping mechanisms. To confirm the hypothesis and identify the origin of the trapping processes, we performed high voltage C-V measurement, individually probing both the gate-to-drain capacitance and the source-to-drain capacitance. All C-V curves were measured at 1 MHz at room temperature (RT) and 150 °C, on transistors with W=69mm. C-V characteristics were measured up to different maximum drain voltage $V_{D,MAX}$ (from 10 V up to 200 V) and the drain bias is swept from 0 V to $V_{D,MAX}$ (positive-going) and subsequently, without removing the bias, from $V_{D,MAX}$ to 0 V (negative going) to see the effect of the field plates (FP). From Figure 4, we can observe the three field plates, highlighted in the cross-sectional schematic of Figure 1, acting in the $C_{GD}$-V curve (solid black line). As $V_{DS}$ increases, the field plates deplete the channel electrons underneath. As soon as the $V_{DS}$ becomes high enough to deplete all channel electrons under one field plate, an increased separation is caused between the gate and the drain, resulting in a sudden drop in $C_{GD}$ [19]. Furthermore, we plot the difference between the positive-going and negative-going bias sweeps to extract the overall variation in trapped charge (Fig. 4). The capacitance depends on the electron concentration in the channel, therefore (i) if in the backward sweep the capacitance is lower we have a stronger depletion indicative of negative charge trapping; on the other hand, (ii) if the capacitance is higher we have less electron trapping or positive charge trapping (Fig. 3).

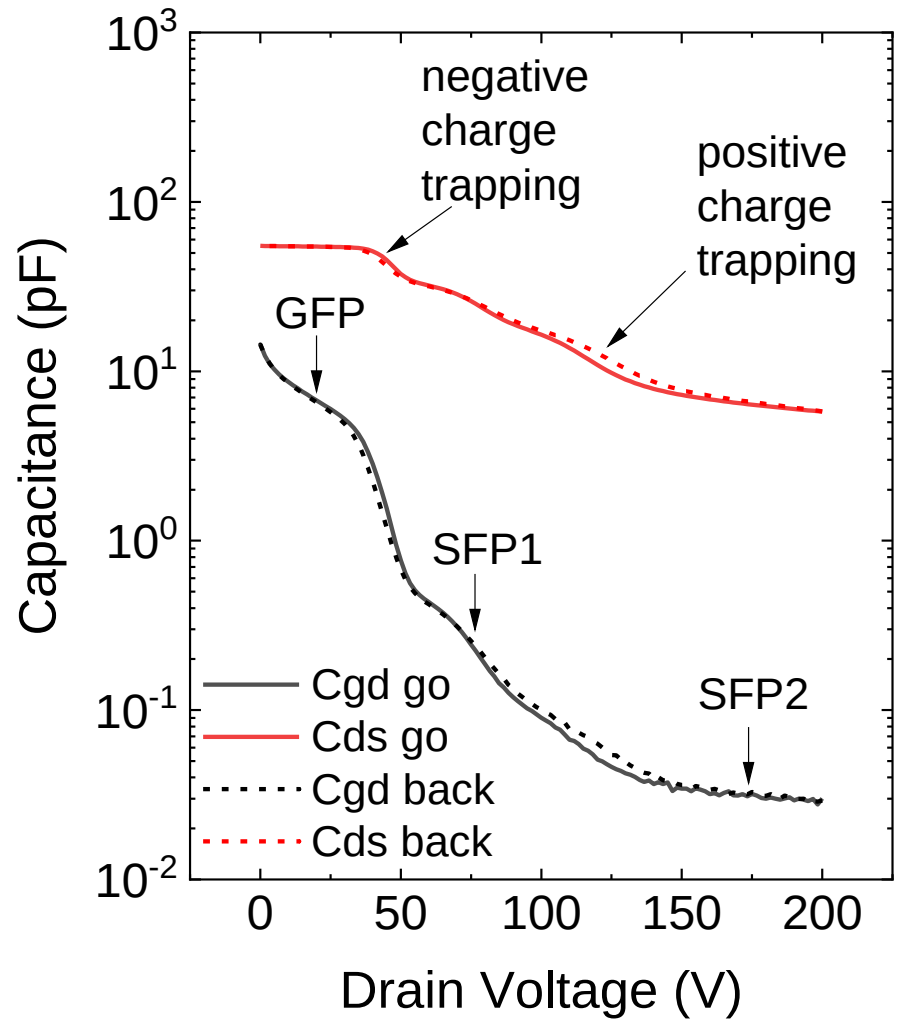

Fig. 3. C-V measurement go and back (to see the FPs acting and their complete pinch-off) up to different drain stop voltages at RT. The timing of a single sweep changes from a minimum of 24 mV/s to a maximum of 472 mV/s.

We found a clear hysteresis for both C-V measurements between the positive-going and the negative going bias sweeps, highlighting difference in charge storage. $C_{GD}$ is more sensitive to trapping under the gate, while $C_{DS}$ is more sensitive to trapping in the buffer. $C_{GD}$ shows a predominant negative and increasing hysteresis for low voltages, as illustrated in Fig. 4 (a), that indicates a negative charge trapping. For high voltages (beyond 50 V), $C_{GD}$ hysteresis exhibits a significantly weakened bias dependence, that agrees very well with the dependence of the $V_{TH}$ variation on the drain voltages (Fig. 2 (b)). This indicates that the negative charge is located under the gate and it suggests that the negative charge storage responsible for $V_{TH}$ shift could take place at the gate stack (**mechanisms 3**, see also [12]) and/or inside the buffer or in the AlGaN barrier [13].

As shown in Fig. 4 (b), the $C_{DS}$-V characteristic presents a cross over point so there is both negative and positive charge trapping; the positive charge trapping starts for high stop drain voltages (> 100V). A possible hypothesis to explain this effect is that at low voltage buffer acceptors are ionized (negative charge, **mechanism 1**), whereas for high voltages (>100 V) increased leakage and/or band-to-band tunneling [14, 15] lead to the formation of positive charge at the buffer/SRL interface (**mechanism 2**). In the backward sweep the positive charge starts to be released.

### *C. Substrate ramps*

Since C-V measurements are sensitive both to trapping at the surface and in the buffer, we performed substrate ramp measurements on TLM structures to assess the trapping properties of the buffer [15, 16, 20-22].

During the measurement the substrate is swept from 0 V down to -200 V, -400 V, -600 V and -800 V, the drain terminal is biased at 1 V, the source terminal is grounded and the drain current is measured. By ramping the substrate to a high (negative) potential, mimicking the off-state operation under the drain contact in a transistor, and monitoring the 2DEG current, changes in the electric field close to the 2DEG are observed. Any charge redistribution in the buffer upon reverse bias will change the electric field and will be sensed as a change in the 2DEG current. As such, the buffer charge trapping or storage will be visible in the substrate ramp characteristics.

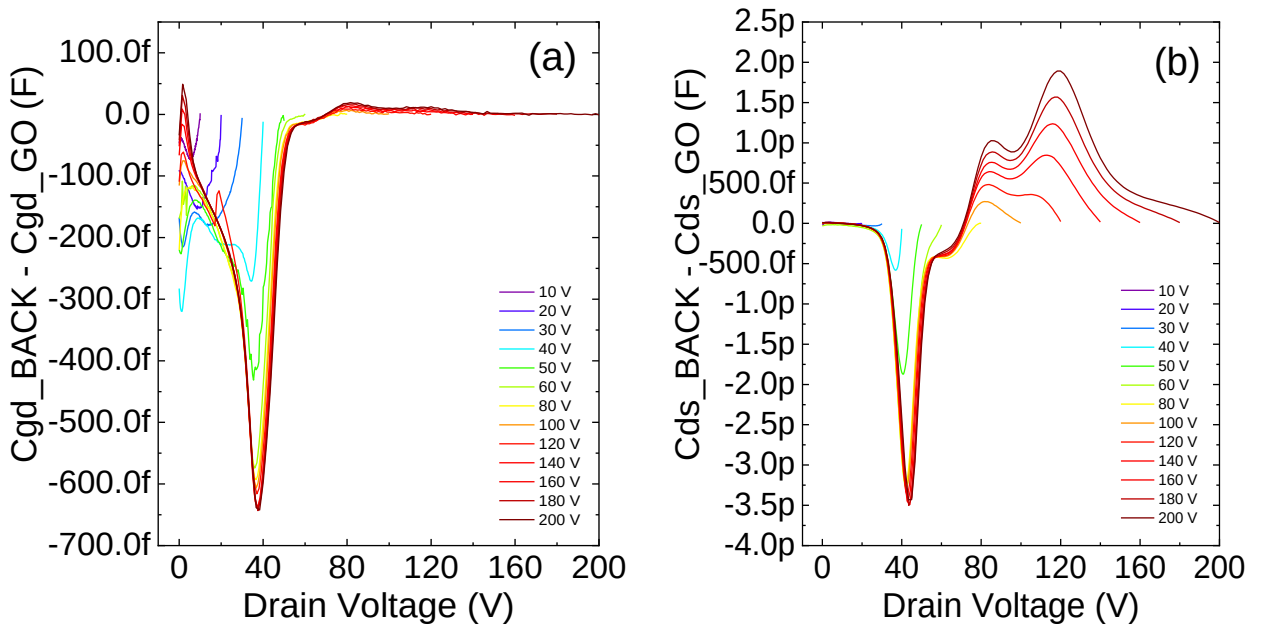

Fig. 4. Difference between the negative-going and positive-going bias sweeps to extract the overall amount of charge for $C_{GD}$ (a) and $C_{DS}$ (b).

The back-biased TLM I-V characteristics are illustrated in Figure 5. Upon the forward sweep, at moderate voltages down to -250 V the current decreases linearly with substrate bias and no charge storage occurs [22]. However, between -250 V and -400 V, the current saturates, i.e. does not change with the substrate voltage: at -250V, the UID GaN channel layer starts to conduct current through a band to band tunneling process, initiated along dislocations. The net effect is positive charge storage at the buffer/SRL interface (**mechanism 2**). This positive charge will reduce the UID GaN electric field resulting in the saturation observed [15, 22]. As the field increases further from -400 V onward, we enter the regime where the leakage starts to occur through the entire buffer structure [15, 22] and the 2DEG current start to decrease again. At high bias (from -550 V onward), the saturation observed is consistent with full depletion of the 2DEG.

On the return sweep, the reverse characteristic is mostly situated above the forward characteristic, indicative of positive charge storage (the current is higher in the reverse than in the forward sweep, indicating a higher 2DEG density) [22]; between -550 V and -450 V upon the return sweep, the reverse curve is slightly below the forward curve, which can be explained by depletion of the 2DEG due to negative stored buffer charge [22]. So the results demonstrate a dominant positive charge variation in the buffer.

As a consequence, the negative charge trapping responsible for $V_{TH}$ and $R_{ON}$ instability is not supposed to be in the buffer, but possibly to trapping in the gate stack or at the surface or in the active region [12].

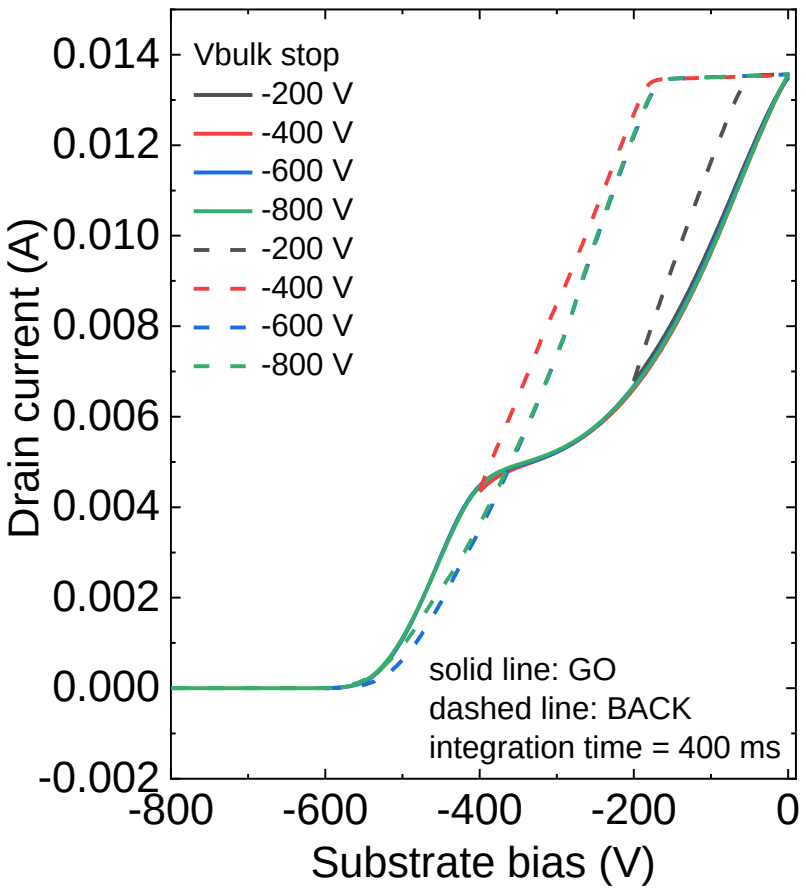

Fig. 5. Substrate ramp up to -200 V, -400 V, -600 V and -800 V on TLM structures.

We repeated the substrate ramp down to -200 V with a wait time (300 s) between each measurement during which we apply UV light (λ = 365 nm) on the floating device under test. We decided to apply UV light during the rest time, before carrying out the next forward and return sweep, in order to remove any memory effect induced by the previous measurement. As illustrated in Figure 6, the results (red curves) show not only a recovery in the drain current but also a negative charge trapping, since the reverse characteristic is situated below the forward characteristic. A possible explanation of this phenomenon is that without the UV light the huge positive charge accumulated at the buffer/SRL interface can compensate the buffer acceptors that are ionized, so **mechanism 2** is dominating. On the other hand, the UV light ionizes most of the buffer acceptors, as a consequence the positive charge storage does not play anymore a strong role, leading to the predominance of the fixed charge of the ionized carbon acceptors in the buffer (**mechanism 1**), which causes the net negative charge variation highlighted by the hysteresis in the results.

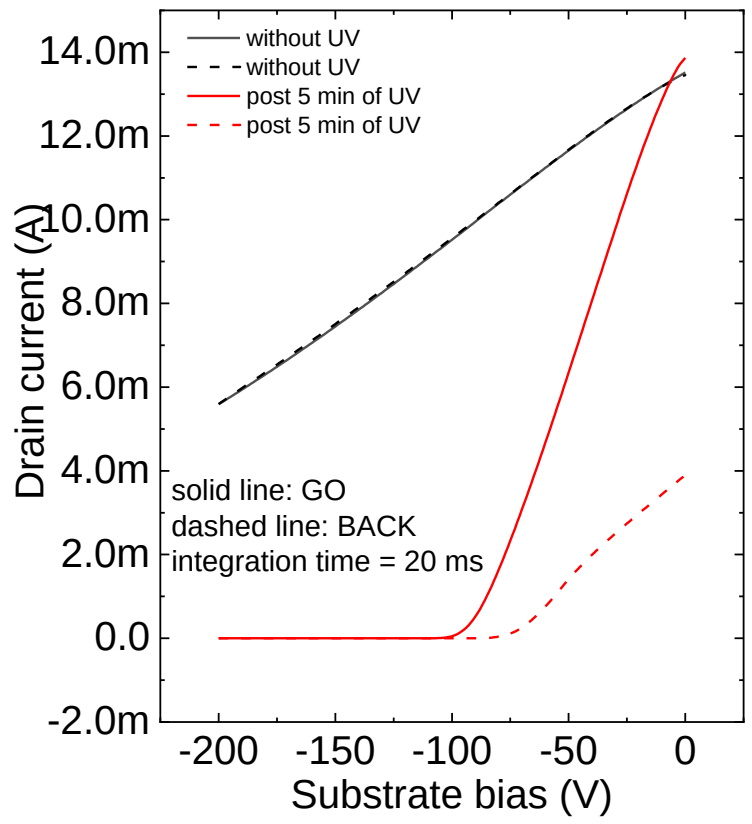

Fig. 6. Substrate ramp down to -200 V with a waiting time (300 s) during which we apply UV light (λ = 365 nm).

Moreover, the pulsed I-V measurements in off-state condition (Fig. 7) show an initial dynamic increase in $R_{ON}$ up to 100 V, then a recovery is observed to the initial value and finally it remains stable. We assume that the trapping in the pulsed measurements is at the gate and it takes place for relatively low voltages up to 100 V and it might impact on the overall $R_{ON}$ [12]. The buffer trapping is not even starting at 100 V but just at higher voltages, is mostly related to positive charge storage, and does not affect the measurements with the times used here.

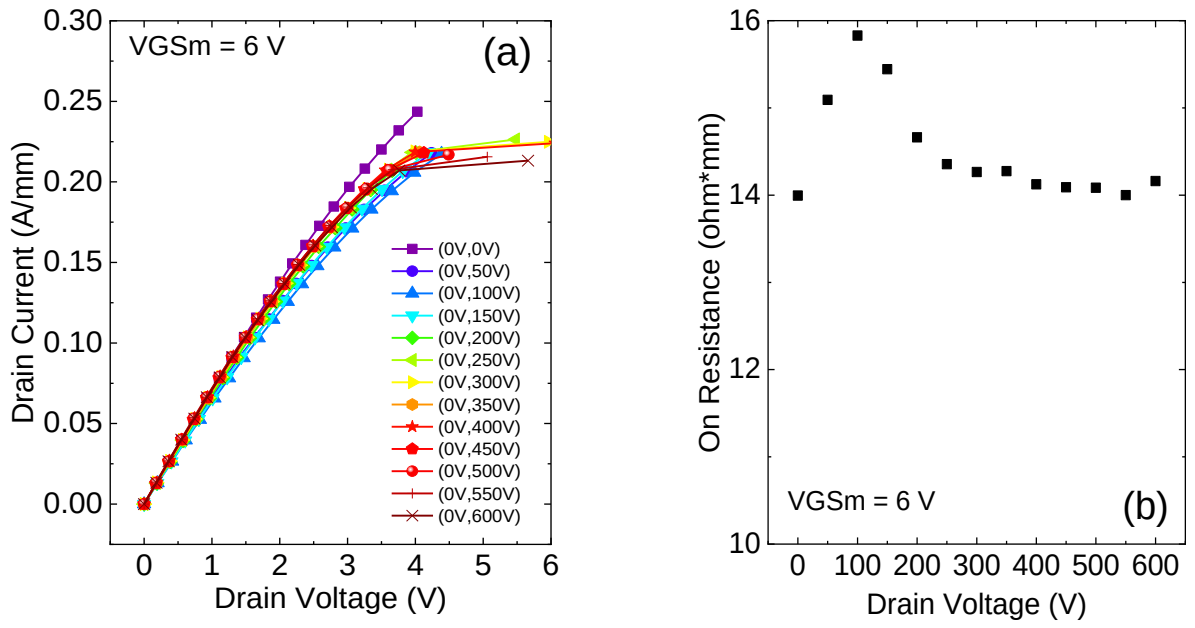

Fig. 7. Pulsed I-V measurements in off-state conditions up to 600 V: (a) $I_DV_D$ characteristic; (b) on resistance as a function of off-state drain stress.

## *D. Pulsed drain current transient*

In order to extrapolate the activation energy of the trapping mechanisms described above, we carried out pulsed drain current transient measurements. The measurement procedure is divided into stress and recovery phase. For the stress phase the device is kept in two different trapping condition ($V_{Gqb}$, $V_{Dqb}$) = (0 V, 45 V), and (0 V, 125 V) for 100s, and the transient investigation is carried out by rapidly changing the gate and drain bias by means of a voltage pulse 10µs long that switches the terminals from the off-state to the on-state ($V_{Gpulse}$ = 6 V, $V_{Dpulse}$ = 1 V).The drain current response is recorded with an inspection timing of 3 samples per decade from 10µs to 100s; then the recovery phase is immediately carried out by keeping the device in rest conditions ($V_{Gqb}$, $V_{Dqb}$) = (0 V, 0 V) and the drain current is periodically recorded as in the stress phase. Several temperatures were tested, from 30 °C to 150°C, to obtain the emission time constant of the deep level at different temperatures. We decided to stress the device at 45 V and 125 V considering the results of the $C_{DS}$-V characteristic and its hysteresis illustrated in Fig. 4 (b): at 45 V we can probe better the negative charge trapping, whereas at 125 V see a stronger effect of the positive charge trapping.

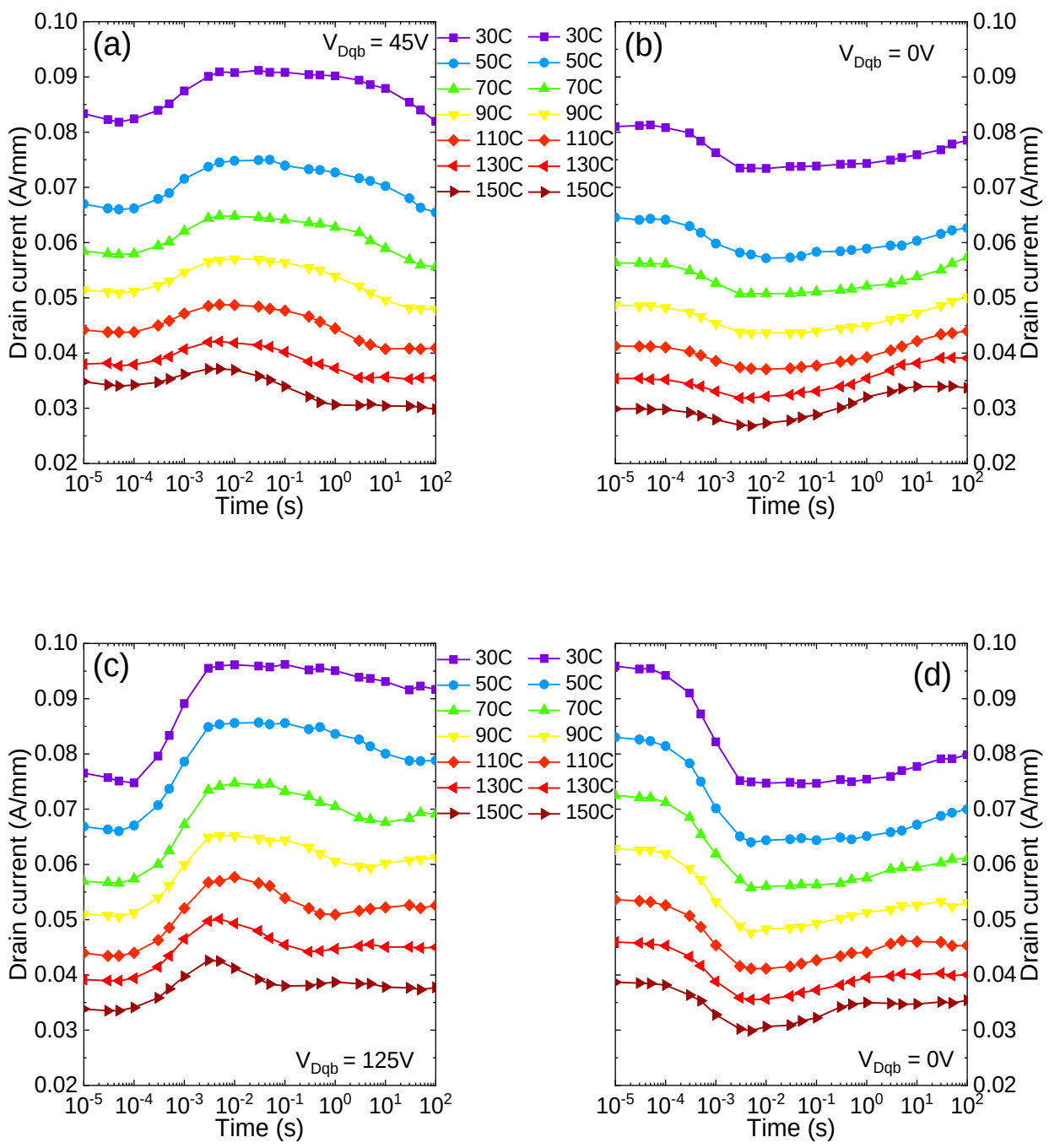

Fig. 8. (a) Pulsed drain current transient for different temperatures (30 °C - 150°C) at 45 V (stress phase); (b) at 0 V recovery phase after stress at 45 V; (c) at 125 V (stress phase); (d) at 0 V recovery phase after stress at 125 V.

The results (Fig. 8) confirm the interplay of two trapping processes responsible for buffer trapping. The drain current transient during the stress phase at 45 V, room temperature has a non-monotonic variation (Fig. 8 (a)): initially, there is a positive charge storage, subsequently followed by the electron trapping, which requires longer time. These two mechanisms take place with different time constants at different temperatures, and at higher temperature the electron trapping becomes stronger with respect to the positive charge storage, possibly due to the high activation energy of the $C_N$ acceptors. The recovery transient (Fig. 8 (b)) shows that after 100s the device is almost back to the initial value.

At 125 V (Fig. 8 (c) and (d)) the positive charge storage is stronger since a much higher vertical electrical field is applied, thus favoring vertical leakage and band to band tunneling through the UID layer. On the other hand, at 45 V (Fig. 8 (a) and (b)) the effects of the two trapping processes are comparable. This confirms the hypothesis that for higher voltages both trapping mechanisms are present but the positive charge accumulation is dominating.

Concerning the filling phase (Fig. 8 (a) and (c), and Fig. 9 (a) and (c)), the positive charge storage in the buffer should depend mostly on the vertical leakage that might be conditioned by a band to band tunneling or other tunneling processes that are not strongly thermally activated [14, 15], leading to the fast and not thermally-activated current increase detected. On the other hand, the negative charge storage is slower and has a strong thermal activation. Moreover, the higher trapping voltage causes a stronger positive charge variation and almost the same level of negative charge trapping, consistently with the findings in Fig. 4.

Concerning the recovery phase (Fig. 8 (b) and (d), and Fig.9 (b) and (d)), the initial current decrease is caused by the redistribution of the accumulated holes in the buffer. Such charges can be located either in a 2-DHG or at shallow traps [14, 15]. On the other hand, the negative charge release leading to the subsequent slow current increase is thermally activated, since it involves by multi-phonon emission from a deep level ($C_N$) [23].

## *E. Activation energy of the trapping mechanisms*

The normalized experimental data of the pulsed drain current transients are fitted by using a stretched exponential function [24] (Figure 9 (a)-(d)) in order to obtain the trapping and de-trapping time constant of the negative charge storage and, therefore, construct the Arrhenius plot to extrapolate the activation energy (Fig. 10).

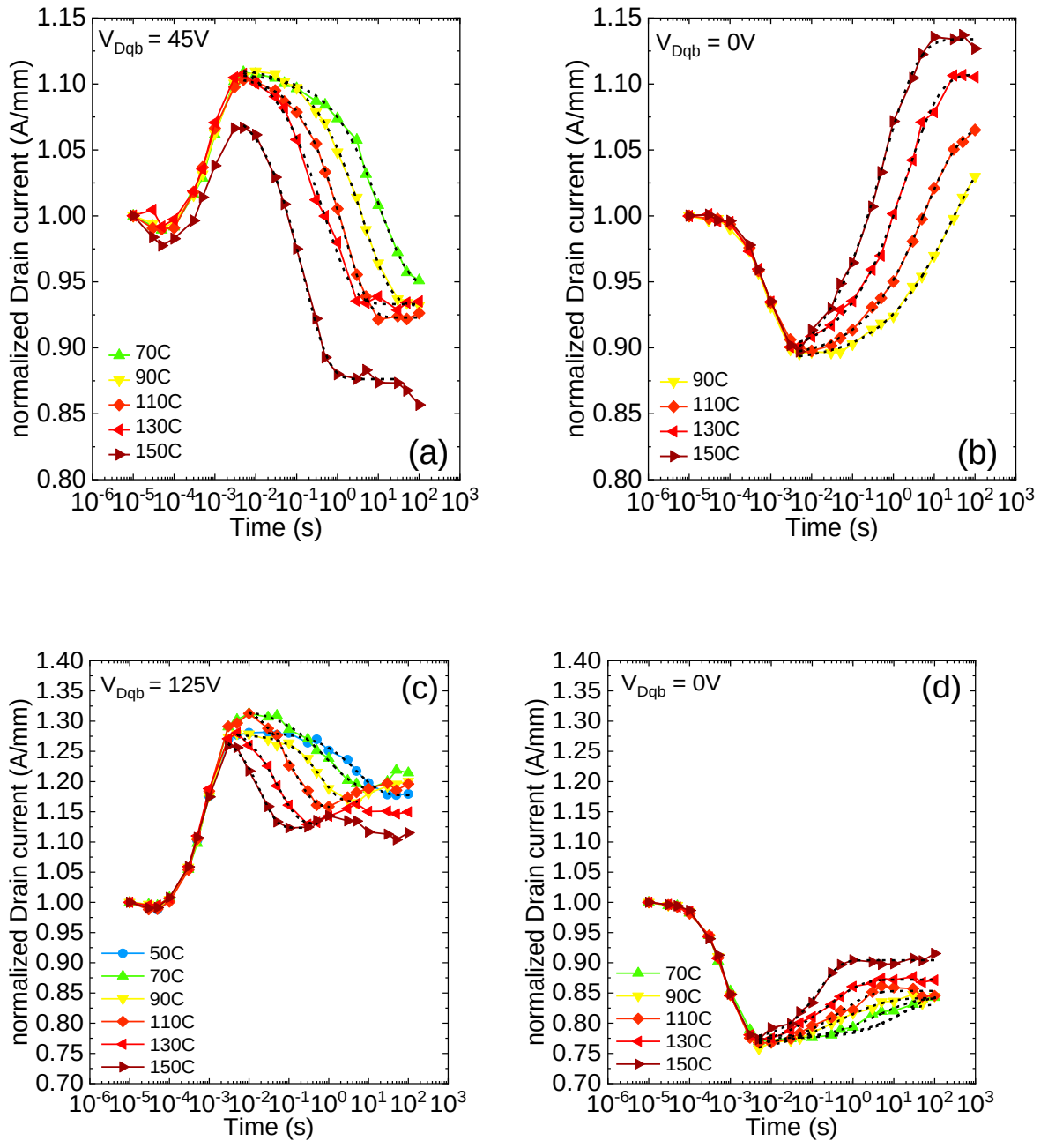

Fig. 9. (a) Normalized pulsed drain current transient for different temperatures and its stretched exponential fit at 45 V (stress phase); (b) at 0 V recovery phase after stress at 45 V; (c) at 125 V (stress phase); (d) at 0 V recovery phase after stress at 125 V.

The extrapolated values of the activation energy for the trapping mechanism at 45 V and at 125 V are 0.6 eV and 0.58 eV, respectively. Recovery transients reveal a de-trapping activation-energy of 0.773 eV after the stress at 45 V and 0.406 eV after the stress at 125 V.

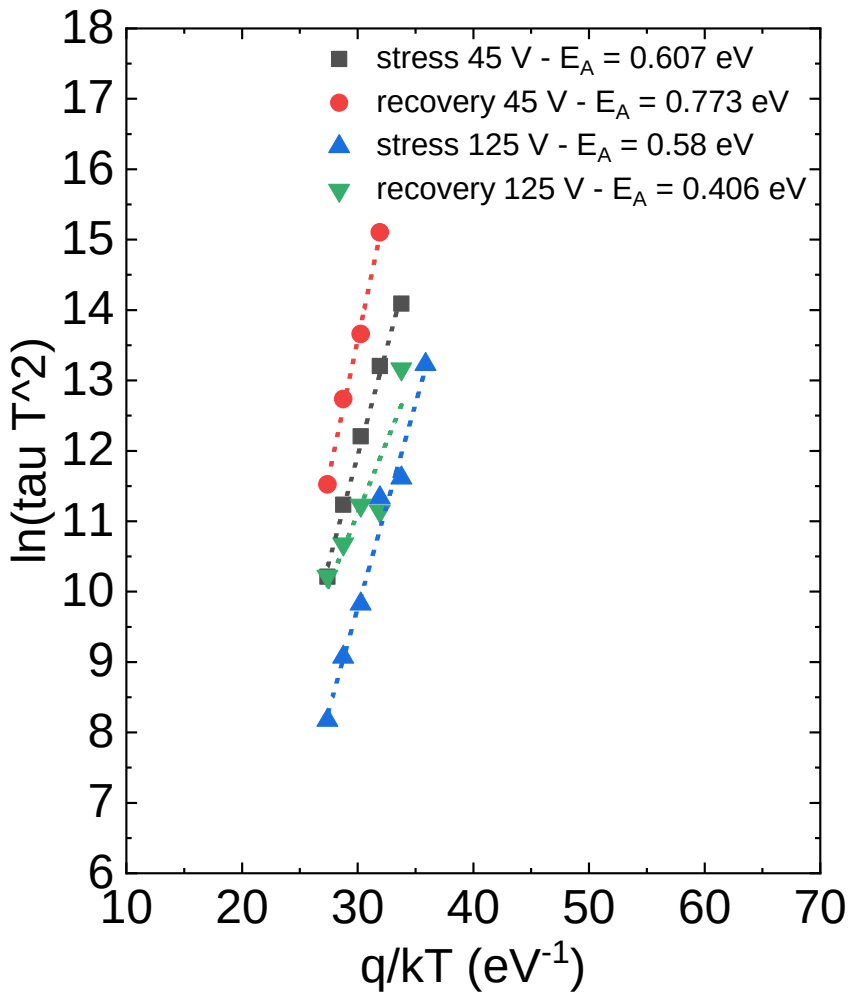

Fig. 10. Arrhenius plot showing that the activation energy $E_A$ of the trapping and de-trapping processes of the negative charge storage, shown in Fig. 9, are indeed very similar, and the $E_A$ found is less than 0.8 eV possibly due to high field.

### F. $V_{TH}$ transient measurements

We performed threshold voltage transient measurements in off-state drain conditions (from 0 V up to 200 V) in order to verify that $V_{TH}$ shift and $R_{ON}$ instabilities have different origins. In a wide time window, from 10 µs to 100 s, devices are kept under stress to induce trapping and fast (10 µs) $I_DV_G$ measurements are repeatedly carried out. The inspection timing is 3 samples per decade from 10µs to 100s. The fast $I_DV_G$ characteristics for some inspection timing are reported in Fig. 11 (a) during zero bias stress and (b) during 50 V stress. The results highlight that there is a positive $V_{TH}$ shift for very short trapping time, followed by a negative shift for both stress conditions, even at 0 V, so we conclude that the $R_{ON}$ variation is independent of the $V_{TH}$ shift (compare the results obtained around threshold and at $V_{GS}$=5 V in Figure 11 (b)).

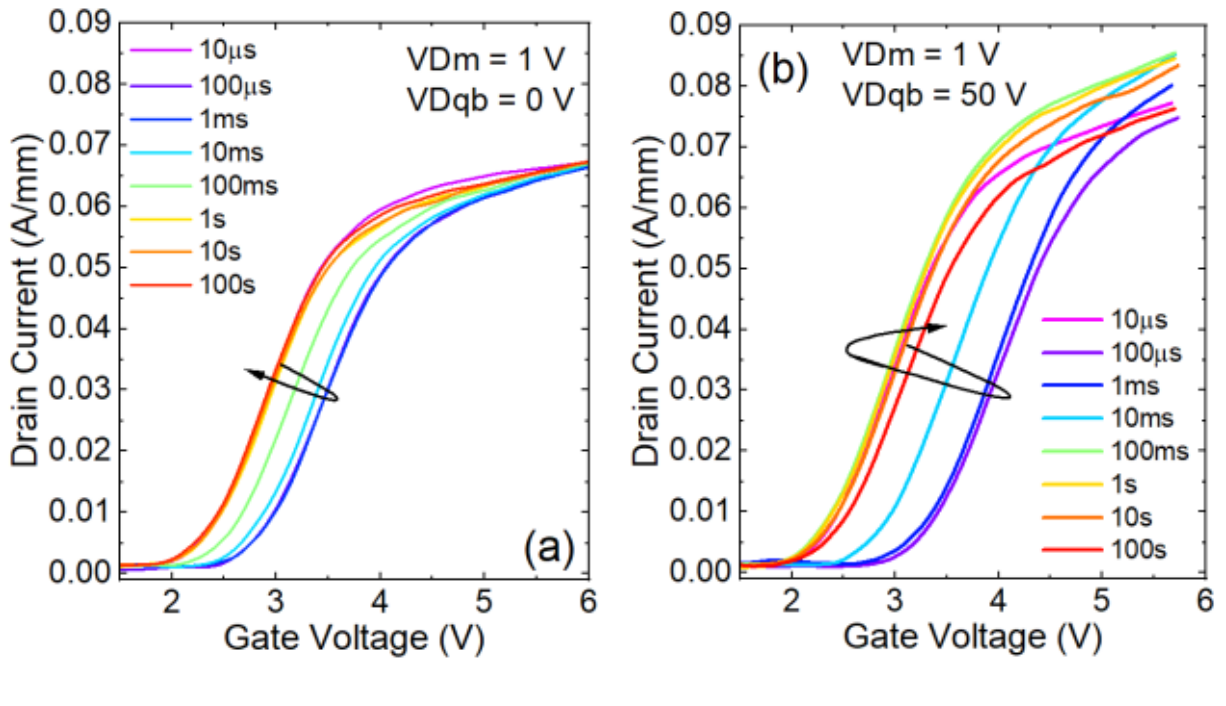

Fig. 11. Fast $I_DV_G$ characteristics carried out from 10 µs to 100 s during OFF-state drain stress at $V_{D,QB}$ = 0 V (a) and $V_{D,QB}$ = 50 V (b).

## IV. Conclusion

We demonstrate and experimentally analyze three charge trapping mechanisms in GaN power HEMTs: 1) the ionization of buffer acceptors, 2) the accumulation of positive charge in the buffer, responsible for a non-monotonic trend of $R_{ON}$, and 3) the storage of negative charge in the gate stack, responsible for a positive $V_{TH}$ shift. The impact of these processes on the capacitance-voltage characteristics and on the turn-on transients is experimentally analyzed, in order to validate the related hypotheses.

For the first time we investigate the time-dependence of the trapping and de-trapping processes described above, finding activation energies ~0.60 eV for the trapping of negative charge storage, and 0.40 eV and 0.77 eV for the recovery of the negative charge. Mechanism 2 (redistribution of holes in the buffer), was found not to be thermally activated, consistently with the hypothesis of charge redistribution in the buffer.

## Acknowledgment

This work was partially supported by the project InRel-NPower (Innovative Reliable Nitride based Power Devices and Applications). This project has received funding from the European Union's Horizon 2020 research and innovation program under grant agreement No. 720527.

## References

[1] N. Ikeda et al., "GaN power transistors on Si substrates for switching applications," Proc. IEEE, vol. 98, no. 7, pp. 1151–1161, July 2010.

[2] M. Kuzuhara and H. Tokuda, "Low-loss and high-voltage III-nitride transistors for power switching applications," IEEE Transactions on Electron Devices, vol. 62, no. 2, pp. 405–413, February 2015.

[3] R. Rupp, T. Laska, O. Häberlen, and M. Treu, "Application specific trade-offs for WBG SiC, GaN and high end Si power switch technologies," in IEDM Tech. Dig., San Francisco, CA, USA, Dec. 2014, pp. 28-31, doi: 10.1109/IEDM.2014.7046965.

[4] Y. Wu, M. Jacob-Mitos, M. L. Moore, and S. Heikman, "A 97.8% efficient GaN HEMT boost converter with 300-W output power at 1 MHz," IEEE Electron Device Lett., vol. 29, no. 8, pp. 824-826, Aug. 2008, doi: 10.1109/LED.2008.2000921.

[5] K. J. Chen, O. Haberlen, A. Lidow, C. L. Tsai, T. Ueda, Y. Uemoto, and Y. Wu, "GaN-on-Si power technology: devices and applications," IEEE Transactions on Electron Devices, vol. 64, no. 3, pp. 779–795, March 2017.

[6] B. Bakeroot, A. Stockman, N. Posthuma, S. Stoffels, and S. Decoutere, "Analytical model for the threshold voltage of *p* -(Al)GaN high electronmobility transistors," *IEEE Trans. Electron Devices*, vol. 65, no. 1, pp. 79–86, Jan. 2018.

[7] L. Efthymiou, G. Longobardi, G. Camuso, T. Chien, M. Chen, and F. Udrea, "On the physical operation and optimization of the p-GaN gate in normally-off GaN HEMT devices," *Appl. Phys. Lett.*, vol. 110, no. 12, Mar. 2017, Art. no. 123502.

[8] I. Hwang, J. Kim, H. S. Choi, H. Choi, J. Lee, K. Y. Kim, J.-B. Park, J. C. Lee, J. Ha, J. Oh, J. Shin, and U.-I. Chung, "p-GaN gate HEMTs with tungsten gate metal for high threshold voltage and low gate current," *IEEE Electron Device Lett.*, vol. 34, no. 2, pp. 202–204, Feb. 2013.

[9] I. Hwang, H. Choi, J. Lee, H. S. Choi, J. Kim, J. Ha, C. Um, S. Hwang, J. Oh, J. Kim, J. K. Shin, Y. Park, U. Chung, I. Yoo, and K. Kim, "1.6kV, 2.9 mΩ cm2 normally-off p-GaN HEMT device," in Proc. ISPSD, Bruges, Belgium, Jun. 2012, pp. 41-44, doi: 10.1109/ISPSD.2012.6229018.

[10] M. H. Kwan, K. Y. Wong, Y. S. Lin, F. W. Yao, M. W. Tsai, Y. C. Chang, P. C. Chen, R. Y. Su, C. H. Wu, J. L. Yu, F. J. Yang, G. P. Lansbergen, H. Y. Wu, M. C. Lin, C. B. Wu, Y. A. Lai, C. W. Hsiung, P. C. Liu, H. C. Chiu, C.

M. Chen, C. Y. Yu, H. S. Lin, M. H. Chang, S. P. Wang, L. C. Chen, J. L. Tsai, H. C. Tuan, and A. Kalnitsky, “CMOS-compatible GaN-on-Si field-effect transistors for high voltage power applications,” in IEDM Tech. Dig., San Francisco, CA, USA, Dec. 2014, pp. 450-453, doi: 10.1109/IEDM.2014.7047073.

[11] Y. Uemoto, M. Hikita, H. Ueno, H. Matsuo, H. Ishida, M. Yanagihara, T. Ueda, T. Tanaka, and D. Ueda, “A normally-off AlGaN/GaN transistor with RonA=2.6mΩcm2 and BVds=640V using conductivity modulation,”in IEDM Tech. Dig., San Francisco, Dec. 2006, pp. 907-910, doi: 10.1109/IEDM.2006.346930.

[12] J. Wei *et al.*, "Charge Storage Mechanism of Drain Induced Dynamic Threshold Voltage Shift in p -GaN Gate HEMTs," in *IEEE Electron Device Letters*, vol. 40, no. 4, pp. 526-529, April 2019, doi: 10.1109/LED.2019.2900154.

[13] L. Efthymiou, K. Murukesan, G. Longobardi, F. Udrea, A. Shibib and K. Terrill, "Understanding the Threshold Voltage Instability During OFF-State Stress in p-GaN HEMTs," in *IEEE Electron Device Letters*, vol. 40, no. 8, pp. 1253-1256, Aug. 2019, doi: 10.1109/LED.2019.2925776.

[14] M. Meneghini, A. Tajalli, P. Moens, A. Banerjee, E. Zanoni, G. Meneghesso,"Trapping phenomena and degradation mechanisms in GaN-based power HEMTs", Materials Science in Semiconductor Processing, Volume 78, 2018, Pages 118-126, ISSN 1369-8001.

[15] M. Uren *et al.*, "“Leaky Dielectric” Model for the Suppression of Dynamic RON in Carbon-Doped AlGaN/GaN HEMTs," in *IEEE Transactions on Electron Devices*, vol. 64, no. 7, pp. 2826-2834, July 2017, doi: 10.1109/TED.2017.2706090.

[16] P. Moens *et al.*, "On the impact of carbon-doping on the dynamic Ron and off-state leakage current of 650V GaN power devices," *2015 IEEE 27th International Symposium on Power Semiconductor Devices & IC's (ISPSD)*, Hong Kong, 2015, pp. 37-40, doi: 10.1109/ISPSD.2015.7123383.

[17] D. Bisi *et al.*, "Kinetics of Buffer-Related RON-Increase in GaN-on-Silicon MIS-HEMTs," in *IEEE Electron Device Letters*, vol. 35, no. 10, pp. 1004-1006, Oct. 2014, doi: 10.1109/LED.2014.2344439.

[18] Arianna Nardo, Matteo Meneghini, Alessandro Barbato, Carlo De Santi, Gaudenzio Meneghesso, Enrico Zanoni, Sebastien Sicre, Luca Sayadi, Gerhard Prechtl, and Gilberto Curatola, “Storage and release of buffer charge in GaN-on-Si HEMTs investigated by transient measurements”, 2020 Appl. Phys. Express 13 074003, doi: https://doi.org/10.35848/1882-0786/ab9623.

[19] R. Chu, A. Corrion, M. Chen, R. Li, D. Wong, D. Zehnder, B, Hughes, and K. Boutros, “1200-V normally off GaN-on-Si field-effect transistors with low dynamic on-resistance,” IEEE Electron Device Lett., vol. 32, no. 5, pp. 632–634, May 2011. doi: 10.1109/LED.2011.

[20] M. Uren et al, “Intentionally Carbon Doped AlGaN/GaN HEMTs: Necessity for Vertical Leakage Paths”, Electron Device Letters vol 35 (3), pp327-329 (2014).

[21] M. Uren et al, “Buffer transport mechanisms in intentionally carbon doped GaN heterojunction field effect transistors”, Applied Physics Letters, vol 104, 263505; doi: 10.1063/1.4885695 (2014).

[22] A. Stockman, M. Uren, A. Tajalli, M. Meneghini, B. Bakeroot and P. Moens, "Temperature dependent substrate trapping in AlGaN/GaN power devices and the impact on dynamic ron," *2017 47th European Solid-State Device Research Conference (ESSDERC)*, Leuven, 2017, pp. 130-133, doi: 10.1109/ESSDERC.2017.8066609

[23] N. Zagni, A. Chini, F. M. Puglisi, P. Pavan, and G. Verzellesi, “The Role of Carbon Doping on Breakdown, Current Collapse and Dynamic On-Resistance Recovery in AlGaN/GaN High Electron Mobility Transistors on Semi-Insulating SiC Substrates,” *Phys. Status Solidi A*, p. 1900762, Dec. 2019. DOI: 10.1002/pssa.201900762.

[24] Rossetto I., Bisi D., De Santi C., Stocco A., Meneghesso G., Zanoni E., Meneghini M., "Performance-Limiting Traps in GaN-Based HEMTs: From Native Defects to Common Impurities", 2019, 10.1007/978-3-319-43199-4_9.